\documentclass[aps,twocolumn,superscriptaddress,floatfix,showpacs,preprintnumbers,amsmath,amssymb]{revtex4}
\usepackage[dvips]{graphicx}
\usepackage{color}
\usepackage{dsfont}
\usepackage{bm}
\usepackage{amsmath}
\usepackage{amsfonts}

\newcommand{\GE}{ |g\rangle \langle e|}
\newcommand{\GG}{ |g\rangle \langle g|}

\newcommand{\EE}{ |e\rangle \langle e|}

\begin{document}
%\nocite{*}
\title{Measurement scheme for the Lamb shift in a superconducting circuit with broadband environment}

\author{V. Gramich}
\affiliation{Institut f\"ur Theoretische Physik, Universit\"at Ulm, Albert-Einstein-Allee 11, 89069 Ulm, Germany}
\author{P. Solinas}
\affiliation{Department of Applied Physics/COMP, Aalto University, P.O. Box 14100, FI-00076 Aalto, Finland}
\affiliation{Low Temperature Laboratory, Aalto University, P.O. Box 13500, FI-00076 Aalto, Finland}
\author{M. M\"ott\"onen}
\affiliation{Department of Applied Physics/COMP, Aalto University, P.O. Box 14100, FI-00076 Aalto, Finland}
\affiliation{Low Temperature Laboratory, Aalto University, P.O. Box 13500, FI-00076 Aalto, Finland}
\author{J. P. Pekola}
\affiliation{Low Temperature Laboratory, Aalto University, P.O. Box 13500, FI-00076 Aalto, Finland}
\author{J. Ankerhold}
\affiliation{Institut f\"ur Theoretische Physik, Universit\"at Ulm, Albert-Einstein-Allee 11, 89069 Ulm, Germany}

\date{\today}

\begin{abstract}
Motivated by recent experiments on quantum mechanical charge pumping in a Cooper pair sluice, we present a measurement scheme for observing shifts of transition frequencies in two-level quantum systems induced by broadband environmental fluctuations. In contrast to quantum optical and related set-ups based on cavities, the impact of a thermal phase reservoir is considered. A thorough analysis of Lamb and Stark shifts within weak-coupling master equations is complemented by non-perturbative results for the model of an exactly solvable harmonic system.
The experimental protocol to measure the Lamb shift in experimentally feasible superconducting circuits is analysed in detail and supported by numerical simulations.
\end{abstract}


\maketitle

\section{Introduction}

Realistic quantum systems are never completely isolated. Even a single atom in zero-temperature vacuum is influenced by the zero-point fluctuations of the electromagnetic field which in turn induces a shift of its transition frequencies known as the Lamb shift~\cite{lamb}. Cavity quantum electrodynamics (QED) provides a particularly convenient set-up to observe this shift since the restricted geometries of the cavities allow the atoms to interact only with the fluctuations of single harmonic fields~\cite{explamb}. Accordingly,  the exploration of fundamental quantum phenomena has reached an unprecedented level of accuracy and has extended our basic understanding of quantum mechanics~\cite{haroche}. Very recently, such a measurement has been performed in a solid state analogue, where a two-level quantum system implemented in the form of a superconducting Cooper pair box (CPB) is embedded in a superconducting waveguide resonator~\cite{wallraff}. The advantage of this device is its tunability which allows to access ranges in the parameter space unavailable in quantum optical set-ups.
%However, the discrete energy levels of the quantum system of interest still interact with a single frequency reservoir the photon number of which is well-controlled.
The theoretical framework here is provided by the Jaynes--Cummings model and its recent extensions (see, e.g.,~\cite{casanova, asshab10}).

In contrast to single-frequency environments, typical reservoirs for mesoscopic solid state devices are characterized by broadband spectral distributions in thermal equilibrium with no convenient access to manipulate the populations of individual modes. Energy exchange between the system and the reservoir as well as other fluctuations lead to dephasing and relaxation, thus limiting time scales over which coherence and entanglement are preserved~\cite{shnirman,breuer,weiss:2008}. An understanding of these processes calls for the well-known theory of {\it{open quantum systems}}, for which several approaches have been developed in the past. Among them, there are formally exact ones such as the path integral representation~\cite{weiss:2008, escher}, and perturbative ones such as Redfield and master equations~\cite{breuer}. Within these weak-coupling formulations even explicit expressions for the reservoir-induced frequency shifts can be derived, while associated experimental observations are still missing.
To fill this gap, in the sequel we discuss and analyse a theoretical proposal to retrieve the Lamb shift and its finite-temperature version (termed also ac-Stark shift according to the shift induced by a finite-amplitude drive)  for a superconducting two-level system embedded in an ohmic environment.

The two-level quantum system studied here is based on the Cooper pair sluice~\cite{niskanen:03}, i.e., a superconducting island separated by two superconducting quantum interference devices (SQUIDs) acting as tunable Josephson junctions. This device is embedded in a superconducting loop together with an additional large Josephson junction employed as a threshold current detector, similar to the Quantronium device~\cite{vion:2002}. By adiabatically tuning the control fields of the sluice (left and right effective Josephson couplings and gate voltage of the island) along a closed path in the parameter space, the accumulated Berry phase~\cite{berry:1984} has been studied experimentally by observing the transferred Cooper pair current as a function of the total phase across the sluice ~\cite{mottonen08}. Furthermore, the impact of the environmental degrees of freedom have been investigated in a set-up where an additional resistor is coupled capacitively to the sluice island~\cite{pekola09,solinasPRB10,environmentpaper,Floquet}. The peculiarity of this architecture is that the temperature of the system is determined by that of the resistor. The same device is analysed here for a spectroscopic measurement; no time-dependent pumping signal is applied. This allows us to determine the Lamb and the Stark shifts for the circuit.

This paper is organized as follows:
In the following section, we briefly recall the weak-coupling approach for open quantum systems to introduce the notation and to give explicit expressions for the reservoir-induced frequency shifts. To validate our results and to gain insight in the regime of stronger system--bath interaction even without the commonly used rotating wave approximation (RWA), the exactly solvable case of harmonic degrees of freedom is discussed in Sec.~\ref{sec:Lambshiftharmonicoscillator}. Section~\ref{sec:experiment} presents the proposed experimental measurement scheme with numerical simulations specified in Sec.~\ref{sec:numerics}. Conclusions are given in Sec.~\ref{sec:conclusions}.

\section{Weak-coupling limit and Lamb shift}\label{sec:microscopicmodel}

%In this section, we summarize briefly the main results for the perturbative treatment of open quantum systems and the Lamb shift in the weak-coupling regime. General expressions are then applied to two-level systems.

\subsection{General theory}\label{general}
The standard microscopic model for open quantum systems \cite{weiss:2008} considers the total Hamilton operator ${H}={H}_S+{H}_R+{H_I}$, composed of a Hamiltonian ${H}_S$ of the system of interest, a reservoir part ${H}_R$ and an interaction operator  ${H_I}$ of the form
\begin{eqnarray}
{H}_R&=&\sum_{n}\,\biggl(\frac{p_n^2}{2 m_n} + \frac{m_n}{2}\omega_n^2 x_n^2\biggr)\,,\nonumber\\
{H_I}&=&-Q\sum_{n}c_n x_n+Q^2\sum_{n}\frac{c_n^2}{2 m_n \omega_n^2}\,.
\label{eq:Hamiltonian}
\end{eqnarray}
Here, the system couples bilinearly to the reservoir via the dimensionless operator $Q$ and the last term in $H_I$, the so-called counter term, guarantees that only the dynamical impact of the reservoir is relevant.
The dynamics of the density operator of the entire system $W(t)$ obeys the Liouville--von--Neumann equation
$\partial_t{W}(t)=-i/\hbar\,[{H},W(t)]$ with an initial state $W(0)$ which factorizes in a system part $\rho(0)$ and the thermal equilibrium of the reservoir $\exp(-\beta H_R)/Z_\beta$ at inverse temperature $\beta=1/k_{\rm B} T$ with partition function $Z_\beta={\rm Tr}_{\rm R}\{\exp(-\beta H_R)\}$.
Hence, the modeling of $H_R$ in terms of harmonic oscillators assumes Gaussian noise properties of the forces $\xi=\sum_n c_n x_n$ of the heat bath with zero mean $\langle \xi(s)\rangle_\beta=0$ and correlation function
\begin{eqnarray}
K(s)&=&\langle \xi(s)\, \xi(0)\rangle_\beta=K'(s)+ i\, K''(s)\nonumber\\
&=&\hbar\int_0^{\infty}\!\frac{d\omega}{\pi} J(\omega)\Biggl[\coth\biggl(\frac{\omega\hbar\beta}{2}\biggr)\cos(\omega\Bigr. \Bigl.s)-i\,\sin(\omega s)\Biggr],\nonumber\\
\label{eq:correlationfunction}
\end{eqnarray}
where the limit to a quasi-continuum of bath modes is performed by introducing a spectral density $J(\omega)=\frac{\pi}{2}\sum_n\frac{c_n^2}{m_n \omega_n}\delta(\omega-\omega_n)$ with dimension of energy.%~\cite{note_generaltheory}.

The relevant operator is the reduced density operator of the system, $\rho(t)={\rm{Tr}_R}\{W(t)\}$, which is obtained by tracing out the reservoir degrees of freedom.
While this procedure does not in general lead to a simple equation of motion for $\rho(t)$, progress is made in the weak-coupling limit upon employing the Born--Markov approximation. This yields the Redfield equation
\begin{equation}
\frac{d\rho(t)}{dt}=-\frac{i}{\hbar}\,\Bigl[{H}_S+Q^2\frac{\gamma(0)}{2},\rho(t)\Bigr]-\mathcal{L}[\rho]\,,
\label{eq:masterequationstandardmodel}
\end{equation}
where the effective impact of the heat bath is captured by
\begin{eqnarray}
\mathcal{L}[\rho]=\frac{1}{\hbar^2}\int_0^{\infty}\!\!\!\!&ds&\!\!  \Bigl\{K'(s)\bigl[Q,[Q(-s),\rho(t)]\bigr]\nonumber\\
& &+i\,K''(s)\bigl[Q,\{Q(-s),\rho(t)\}\bigr]\Bigr\}\,.
\label{eq:Liouvilleoperator}
\end{eqnarray}
Here, the time dependence of $Q(s)$ is taken in the interaction picture with respect to ${H}_S$ and $\{\cdot\,,\cdot\}$ denotes the usual anticommutator. The contribution $\sim\gamma(0)$ in Eq.~(\ref{eq:masterequationstandardmodel}) is related to the counter term in (\ref{eq:Hamiltonian}) with $\gamma(0)=\int d\omega\,[J(\omega)/\pi\omega]$ \cite{weiss:2008}.

For the explicit representation of the operator master equation (\ref{eq:masterequationstandardmodel}) it is convenient to use the eigenstates of the bare system, i.e., ${H}_S|n\rangle=E_n|n\rangle$. In the absence of external time-dependent fields, the conventional procedure is to apply an additional RWA, where off-resonant, fast-oscillating terms are neglected. Thus, diagonal and off-diagonal elements of the density matrix $\rho_{mn}=\langle m|\rho|n\rangle$ decouple to finally yield in the Schr\"odinger picture
\begin{equation}
\frac{d\rho_{mn}}{dt}=-i\,\tilde{\omega}_{mn}\rho_{mn}+\delta_{mn}\sum_{k\neq n}W_{nk}\,\rho_{kk}-\Gamma_{mn}\,\rho_{mn}\,.
\label{eq:RWAequation}
\end{equation}
For the populations $\rho_{nn}(t)$ transitions between eigenstates are captured by rates
$W_{nm}=|\langle m|Q|n\rangle|^2 D(\hbar\omega_{nm})/\hbar^2$, where $\omega_{nm}=(E_n-E_m)/\hbar$ denotes the bare transition frequency and
\begin{equation}
D(\hbar\omega)=2\,{\rm Re}\left\{\int_{0}^{\infty}\!\! ds\, K(s) e^{-i\omega s}\right\}=2\hbar J(\omega) n_{\beta}(\omega)
\label{eq:Dfunction}
\end{equation}
is the spectral noise density function with the single-particle Bose distribution
$n_{\beta}(\omega)=1/[\exp(\omega\hbar\beta)-1]$. The off-diagonal elements of the density matrix decay according to
\begin{equation}
\Gamma_{mn}=\frac{\Gamma_{mm}+\Gamma_{nn}}{2}+\frac{D(0)}{2\hbar^2}\Bigl(\langle m|Q|m\rangle-\langle n|Q|n\rangle\Bigr)^2
\label{eq:decayrates}
\end{equation}
with
$\Gamma_{mm}=\sum_{k\neq m}W_{km}$. The interaction between system and reservoir leads also to a shift in the bare transition frequencies
$\tilde{\omega}_{mn}=\omega_{mn}-\delta\omega_{mn}$ which is given in the Born--Markov limit by
\begin{equation}
\delta\omega_{mn}=\sum_k\biggl[-|\langle m|Q|k\rangle|^2\frac{\tilde{D}(\hbar\omega_{km})}{2\hbar^2}+|\langle n|Q|k\rangle|^2\frac{\tilde{D}(\hbar\omega_{kn})}{2\hbar^2}\biggr],
\label{eq:BornMarkovLambshift}
\end{equation}
where bath fluctuations appear in form of
\begin{eqnarray}
\tilde{D}(\hbar\omega)&=&2\,{\rm Im}\left\{\int_0^{\infty} \!\!ds \, K(s) e^{-i\omega s}\right\}+\hbar\gamma(0)\nonumber\\
&=&-2\, K'_{\rm sin}(\omega)+\hbar\,\omega\,\gamma_{\rm sin}(\omega)\,,
\label{eq:Dtildefunction}
\end{eqnarray}
determined by
\begin{eqnarray}
&&K'_{\rm sin}(\omega)=\int_0^{\infty}\!\! ds\, K'(s) \sin(\omega s)\nonumber\\
&&=\gamma_{\rm sin}(\omega)\frac{\hbar\omega}{2}\, {\rm coth}\left(\frac{\omega\hbar \beta}{2}\right)+\frac{2\omega}{\beta}\sum_{n=1}^\infty\frac{\nu_n \hat{\gamma}(\nu_n)}{\omega^2+\nu_n^2}\,. \nonumber\\
\label{eq:sintransform}
\end{eqnarray}
Above, the sine transformation $\gamma_{\rm sin}$ and the Laplace transformation $\hat{\gamma}$ of the classical damping kernel $\gamma(s)=\int d\omega\,[J(\omega)/\pi\omega]\cos(\omega s)$ appear and $\nu_n=2\pi n/\hbar\beta$ denote the Matsubara frequencies.
The above expression for $\delta\omega_{mn}$ captures the impact of reservoir fluctuations for both finite temperatures (ac Stark shift) and at zero temperature (Lamb shift). Since typical solid-state reservoirs carry a broad range of spectral modes covering low as well as high frequencies, the system is embedded in an environment with a continuum of transition frequencies in contrast to the situation of a single-mode environment such as a cavity. Accordingly, there are always low-frequency excitations present even at low temperatures and thus a clear separation between Stark and Lamb shift is not possible. Hence, in the sequel we simply use the notion Lamb shift for all reservoir-induced shifts of the transition frequencies.

Let us specify the range of validity of the weak-coupling master equation~(\ref{eq:RWAequation}). It applies as long as (i) a typical time for relaxation of the system towards thermal equilibrium
exceeds by far the retardation time of the reservoir and (ii) the level broadening induced by the heat bath is small compared to typical energy gaps of the bare system. Given a relaxation time of order $1/\gamma$ (with a typical coupling strength $\gamma$ to the heat bath), retardation times of the bath of order $1/\omega_c$ (typical cut-off frequency $\omega_c$), and $\hbar\beta$ (thermal time scale) as well as a typical system frequency $\omega_0$, these conditions are fulfilled if
\begin{equation}
\gamma\hbar\beta\ll 1\ \ \mbox{and} \  \  \gamma\ll {\omega_0}\ll {\omega_c}\, .
 \end{equation}
 This allows for very low temperatures compared with system energies such that the quantum optical limit $\gamma/\omega_0\ll 1$ practically includes the zero-temperature limit.

\subsection{Two-level system}\label{TLS}
To illustrate the above results let us consider a two-level system, namely,
\begin{equation}
H_S=\frac{1}{2} E_z\, \sigma_z + \frac{1}{2} E_x\, \sigma_x\, ,
\label{spinH}
\end{equation}
where $\sigma_i$ denote Pauli matrices and $E_i$ are corresponding energies. In the experimental implementation to be discussed below, the eigenstates of $\sigma_z$ are associated with charge states which interact via $Q= \sigma_z$ with a surrounding reservoir so that \cite{note_TLS}
\begin{equation}
H_I=-\sigma_z \sum_n  c_n x_n\, .
\end{equation}

Within the eigenstate representation with basis $\{|g\rangle, |e\rangle\}$  the system Hamiltonian (\ref{spinH}) reads
\begin{equation}
{H}_S=\frac{\hbar\omega_0}{2}(|e\rangle\langle e|-|g\rangle\langle g|)\, ,
 \end{equation}
 where $\hbar\omega_0=\sqrt{E_z^2+E_x^2}$. The operator $Q$ takes the form
 \begin{equation}
 Q=\sin(\eta) (|e\rangle\langle g|+|g\rangle\langle e|)+ \cos(\eta) (|e\rangle\langle e|-|g\rangle\langle g|)
 \end{equation}
 with the mixing angle determined by ${\rm tan}(\eta)=E_x/E_z$. This system is assumed to interact with a reservoir of Drude type
 \begin{equation}
 J(\omega)=\hbar\, \alpha\frac{\omega\omega_c^2}{\omega^2+\omega_c^2}\, ,
 \label{drude}
   \end{equation}
  where $\omega_c\gg \omega_0$ denotes a high cut-off frequency and $\alpha$ is a dimensionless coupling constant related to the typical dissipation strength discussed above via  $\gamma\sim \alpha\omega_0$.

For the Lamb shift to be relevant, it must exceed the level broadening induced by the relaxation and excitation processes which is of order $\hbar \Gamma_{ee}$ [see Eq.~(\ref{eq:RWAequation})]. In the high temperature regime $\omega_0\hbar\beta\ll 1 $, one has $\Gamma_{ee}\sim \alpha/\hbar\beta$ and Eq.~(\ref{eq:BornMarkovLambshift}) yields a Lamb shift that is negligible, namely, $|\delta\omega_{eg}|\sim  (\omega_0/\omega_c)\,  \Gamma_{ee}$.
The reservoir-induced Lamb shift is significant only at low temperatures $\omega_0\hbar\beta>1$, where $\Gamma_{ee}\sim \alpha\omega_0$, while an explicit evaluation of Eq.~(\ref{eq:BornMarkovLambshift}) together with Eqs.~(\ref{eq:Dtildefunction}) and (\ref{eq:sintransform}) provides an effective transition frequency $\tilde{\omega}_{eg}=\omega_0-\delta\omega_{eg}$ with
\begin{eqnarray}
\frac{\delta\omega_{eg}}{\omega_0}=\frac{2 \alpha}{\pi}{\rm Re}\left\{\psi(i\omega_0/\nu_1)-\psi(\omega_c/\nu_1)\right\}\, .
\label{weakLamb}
\end{eqnarray}
Corrections to the above expression are of order $\omega_0/\omega_c$ and $\psi$ denotes the digamma function.
For very low temperatures this result reduces in leading order to
\begin{equation}
\frac{\delta\omega_{eg}}{\omega_0}=-\frac{2\alpha}{\pi}\,{\rm ln} \left( \frac{\omega_c}{\omega_0} \right)+\frac{2\alpha \pi}{3(\omega_0\hbar\beta)^2} \,,
\end{equation}
which exceeds the level broadening by a characteristic logarithmic dependence on the cut-off frequency, also
known from the atomic Lamb shift~\cite{lamb}. For somewhat higher temperatures the Lamb shift (\ref{weakLamb}) decreases compared to the above asymptotic value.

\section{Lamb shift of a damped harmonic oscillator}\label{sec:Lambshiftharmonicoscillator}

In the case of a harmonic system, the reduced quantum dynamics can be calculated {\textit{exactly}} for arbitrary spectral densities, temperatures, coupling strengths, and even including correlated initial states \cite{weiss:2008}. For factorizing initial states, an exact master equation with time-dependent coefficients was first derived by Haake and Reibold~\cite{haakereibold:85, karrlein:1997}. In the case of reservoirs with sufficiently large cut-off frequencies and in the regime $\gamma\hbar\beta<1$, these coefficients become time-independent after a transient period of time short compared to the relaxation dynamics. The corresponding master equation allows to extract analytical expressions for the Lamb shift that can be used  to specify the range of validity of the weak-coupling results of Eqs.~(\ref{eq:BornMarkovLambshift}) and (\ref{weakLamb}). In particular at sufficiently low temperatures the harmonic oscillator reduces to a two-level system~\cite{weiss:2008}. Moreover, it provides insight in the non-perturbative regime of strong coupling which in turn will help us to identify the optimal parameter range to detect the Lamb shift in a broadband solid-state environment.

Hence, we study a system Hamiltonian ${H}_S=p^2/2M + M\omega_0^2 q^2/2$ of an oscillator with mass $M$ which interacts via $Q=q\sqrt{M\omega_0/\hbar}$ with a reservoir of Drude type $J(\omega)= M\gamma \omega\omega_c^2/(\omega^2+\omega_c^2)$ [note that for continuous degrees of freedom it is convenient to use a spectral density which differs from that for discrete systems (\ref{drude}) by a factor $M\omega_0/\hbar$ with $\gamma=\alpha\omega_0$].
 One then obtains after a transient period for the reduced dynamics~\cite{haakereibold:85, karrlein:1997}
\begin{eqnarray}
\dot{\rho}(t)
&=&-\frac{i}{\hbar}[{H}_S,\rho(t)]-\frac{i\gamma}{2\hbar}\Bigl[q,\{p,\rho(t)\}\Bigr]\nonumber\\
&&-\frac{D_p}{\hbar^2}\Bigl[q,[q,\rho(t)]\Bigr]+\frac{D_q}{\hbar^2}\Bigl[p,[q,\rho(t)]\Bigr]
\label{eq:HaakeReiboldohmicdamping}\, .
\end{eqnarray}
The diffusion coefficients take for $\omega_c\gg \omega_0$ \cite{karrlein:1997} the form $D_p=\gamma\langle p^2\rangle$ and $D_q=M\omega_0^2\langle q^2\rangle-\langle p^2\rangle/M$ with the {\it exact} variances in thermal equilibrium \cite{haakereibold:85, weiss:2008}
\begin{eqnarray}
\langle q^2\rangle&=&\frac{1}{M\beta}\sum_{n=-\infty}^{\infty}\frac{1}{\omega_0^2+\nu_n^2+|\nu_n|\hat{\gamma}(|\nu_n|)}\,,\nonumber\\
\langle p^2\rangle&=&\frac{M}{\beta}\sum_{n=-\infty}^{\infty}
\frac{\omega_0^2+|\nu_n|\hat{\gamma}(|\nu_n|)}{\omega_0^2+\nu_n^2+|\nu_n|\hat{\gamma}(|\nu_n|)}\, .
\label{eq:equilibriummoments}
\end{eqnarray}

Upon representing Eq.~(\ref{eq:HaakeReiboldohmicdamping}) in the usual eigenstate basis $\{|n\rangle\}$ of ${H}_S$ and invoking the RWA, one derives in the subspace spanned by $\{|0\rangle, |1\rangle\}$ a transition frequency $\tilde{\omega}_{10}=\omega_0-\delta\omega_{10}$ with
\begin{eqnarray}
\delta\omega_{10}&=&\frac{D_q}{\hbar}\nonumber\\
&=&-\frac{1}{\hbar\beta}
\sum_{n=-\infty}^{\infty}\frac{|\nu_n|\hat{\gamma}(|\nu_n|)}{\omega_0^2+\nu_n^2+|\nu_n|\hat{\gamma}
(|\nu_n|)}\, .
\label{eq:RWAlambshift}
\end{eqnarray}
Apparently, the Lamb shift is directly associated with the deviation from the equipartition relation, i.e., $M\omega_0^2\langle q^2\rangle\neq\langle p^2\rangle/M$, induced by quantum fluctuations in the reservoir. In agreement with the findings of the previous section, it is thus expected to be strong at very low temperatures. For weak dissipation, i.e., $\gamma/\omega_0\ll 1$, the sum in Eq.~(\ref{eq:RWAlambshift}) reduces  in leading order to the sum in Eq.~(\ref{eq:sintransform}) which yields an expression of the form of Eq.~(\ref{weakLamb}) with $\alpha\rightarrow\gamma/\omega_0$. We also note that the relaxation rate $\Gamma_{ee}\sim \gamma[\langle p^2\rangle/(M\hbar\omega_0)+1]$ reduces in the weak-coupling limit to the known result $\Gamma_{ee}\sim \gamma [{\rm coth}( \omega_0\hbar\beta/2)+1]$.

The Lamb shift can also be extracted from Eq.~(\ref{eq:HaakeReiboldohmicdamping}) without applying the RWA. Restricting ourselves again to the subspace spanned by the ground and the first excited state of the oscillator, the corresponding coupled equations for the density matrix elements can be easily solved to provide
\begin{equation}
\frac{\delta\omega_{10}^{\rm{nRWA}}}{\omega_0}=1-\sqrt{1-\frac{2 D_q}{\hbar\omega_0}-\frac{D_p^2}{M^2\hbar^2\omega_0^4}}\,.
\label{eq:nonRWALambshift}
\end{equation}
In the low-temperature range, an expansion for weak coupling gives $\delta\omega^{\rm{nRWA}}_{10}\approx D_q/\hbar+D_q^2/(2\hbar^2\omega_0)+ D_p^2/(2M^2\hbar^2\omega_0^3)$
 so that the RWA Lamb shift in Eq.~(\ref{eq:RWAlambshift}) is obtained in leading order with corrections of order $[(\gamma/\omega_0) {\rm ln}(\omega_c/\omega_0)]^2$ at very low temperatures. Based on the full expressions, one can also show that in the range $\gamma\hbar\beta<1$  the ratio $|\delta\omega^{\rm{nRWA}}|/\Gamma_{ee}$ decreases with increasing dissipation strength $\gamma$. This verifies the assumption that the low damping and high cut-off frequency domain constitutes the optimal detection window.
\begin{figure}
	\begin{center}
	\includegraphics[width=7.5cm]{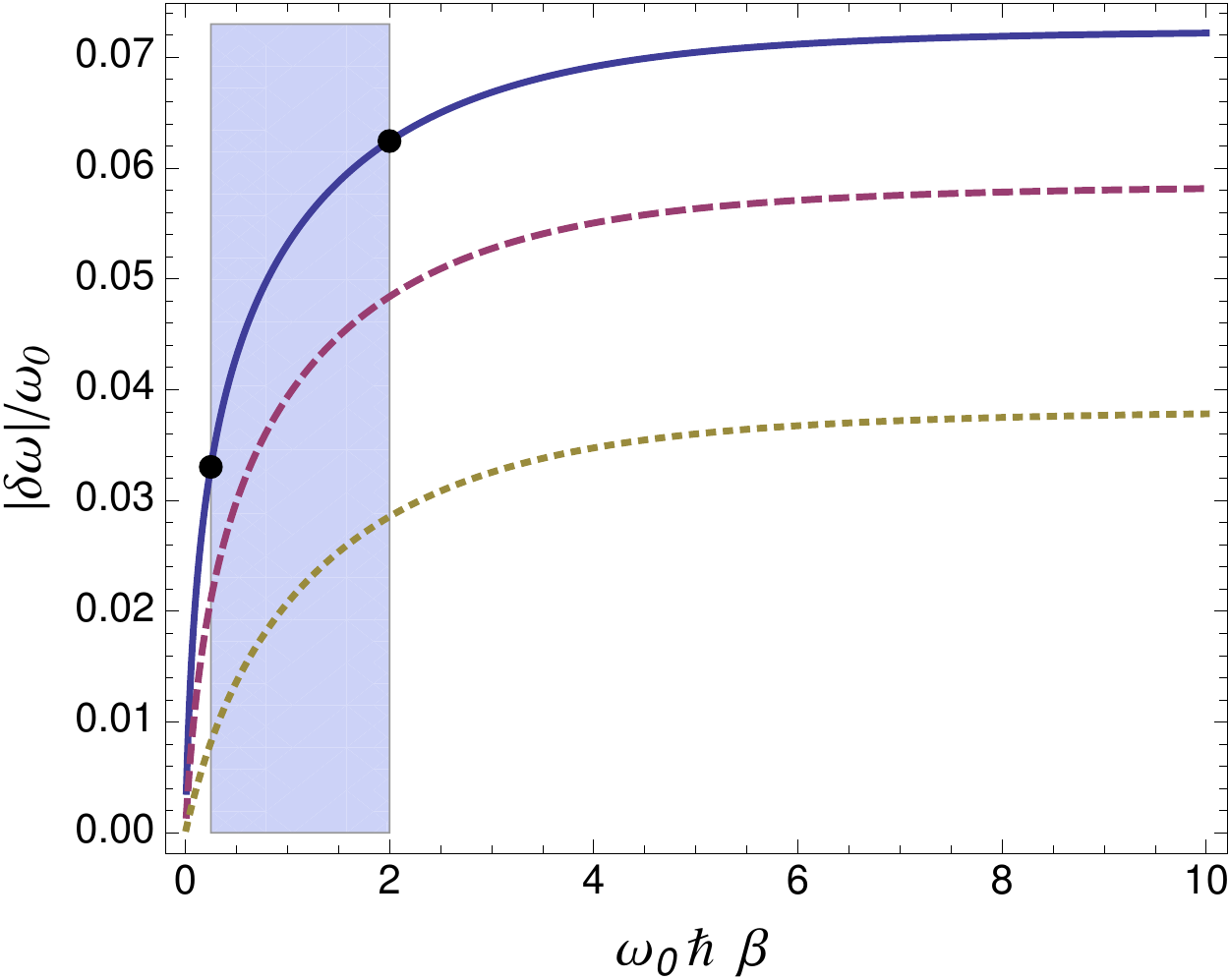}	
\end{center}
	\caption[]{(Colour online) Lamb shift according to  Eq.~(\ref{eq:RWAlambshift}) for  a Drude spectral density as a function of the dimensionless inverse temperature $\omega_0\hbar\beta$. The cut-off frequencies increase from bottom to top with $\omega_c/\omega_0=10$ (dotted line), $\omega_c/\omega_0=40$ (dashed line), and $\omega_c/\omega_0=100$ (solid line). The coupling constant is $\gamma/\omega_0=0.05$. The shaded region and the dots denote the temperature range and the Lamb shifts accessible for the discussed experiment.
}
	\label{fig:lambshiftexact}
\end{figure}

We illustrate the above findings for typical parameters in realistic circuits to be discussed below. Figure~\ref{fig:lambshiftexact} displays the Lamb shift in the relevant weak-coupling case, for which Eq.~(\ref{eq:RWAlambshift}) and the full expression (\ref{eq:nonRWALambshift}) coincide up to minor deviations at low temperature. Whereas the shift depends on the coupling strength, the cut-off frequency, and the temperature, possibly the most convenient way to unambiguously identify the Lamb shift is to detect its temperature dependence. This way, an upward shift of about 6\% of the bare transition frequency drops substantially towards higher temperatures.
The optimal range of parameters for observing the temperature dependence of the shift corresponds to weak friction and large cut-off frequencies.

\section{Experimental proposal
}\label{sec:experiment}
In this section, we discuss a possible way to measure the Lamb shift for a heat bath with a broadband spectral density in a circuit containing a Cooper pair sluice~\cite{niskanen:03}. A brief description of the actual set-up is given in the next section followed by an analysis of the detection proposal.

\subsection{Physical system and its description}

\begin{figure}
    \begin{center}
    \includegraphics[width=7.5cm]{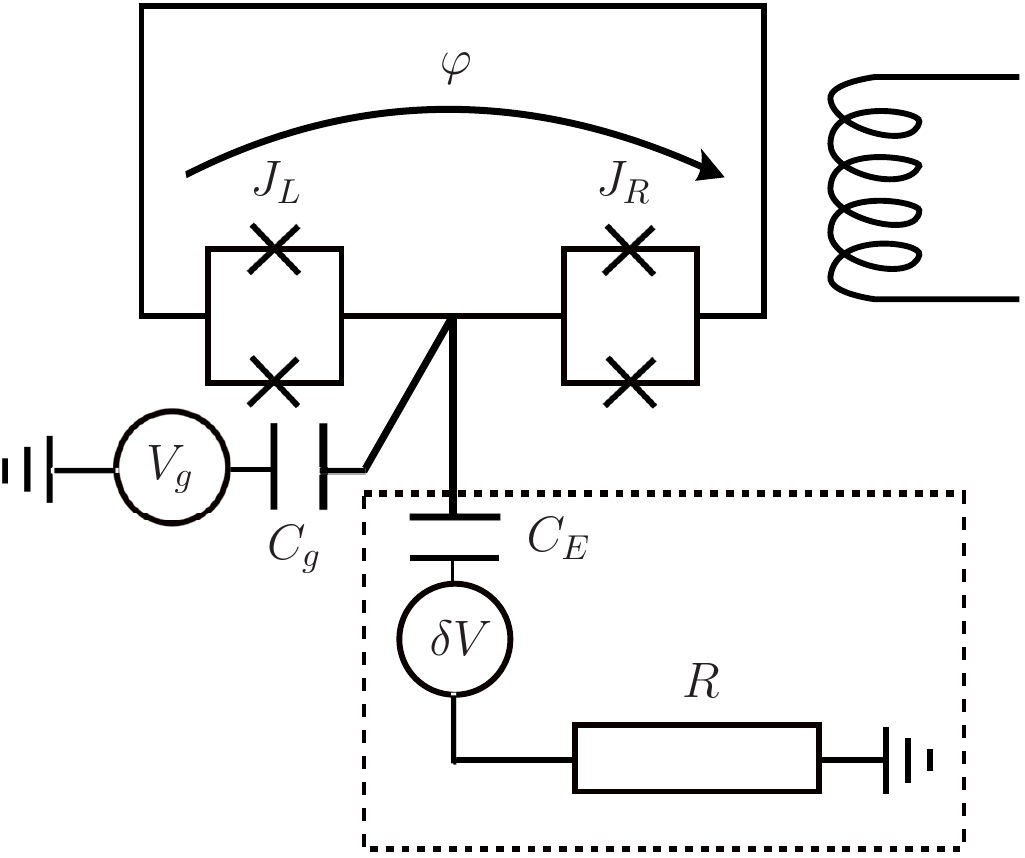}
    \end{center}
    \caption{Circuit diagram for the Cooper pair sluice, the artificial environment (in the dashed box), and the coil used for exciting the system with external drive (on the right).
    }
    \label{fig:system}
\end{figure}

The device we consider here, the sluice, is shown in Fig.~\ref{fig:system}. It consists of a superconducting island separated by two SQUIDs acting effectively as single tunnel junctions with tunable Josephson energies $J_{L,R}$.
The electrostatic potential on the island can be tuned by varying a gate voltage, $V_g$.
The experimental access to the parameters $J_{L,R}$ and $V_g$ allows for a full control of the system and makes it an excellent prototype for detecting fundamental quantum processes.
Note that a similar device has already been used experimentally to study the geometric Berry phase related to adiabatic ground-state pumping of Cooper pairs~\cite{mottonen08,mottonen06}.
The Hamiltonian of the sluice can be expressed as~\cite{pekola09,niskanen:03}
\begin{equation}
{H}_S =E_C(N-N_g)^2 -J_L \cos\left(\frac{\varphi}{2} - \theta\right) -J_R \cos \left(\frac{\varphi}{2} + \theta\right),
\label{eq:H_S}
\end{equation}
where $\varphi=\varphi_R+\varphi_L$ is the superconducting phase difference across the sluice,
$N_g=C_gV_g/2e$ is the normalized gate charge,  $E_{C}= 2e^2/C_{\Sigma}$ is the charging energy of the sluice, $C_g$ is the gate capacitance, and $C_{\Sigma}$ the total capacitance of the island. Due to $[N_k, \varphi_k]=-i$ the Cooper pair number operator $N_k$ of the $k$-th SQUID has a representation
$N_k=-i \partial_{\varphi_k} $ ($k=L, R$), while the number operator of excess Cooper pairs on the island $N$ is related to the corresponding phase difference
$\theta=(\varphi_R-\varphi_L)/2$ by $N= -i \partial_\theta$.
If the device is operated in the charging regime, i.e., $E_C \gg \mbox{max}\{J_L,J_R\}$, and the gate charge is close to a half integer, only the two lowest-lying charge states are important and we can adopt the two-state approximation. Below, states with 0 and 1 excess Cooper pairs on the island are denoted by $|0\rangle$ and $|1\rangle$, respectively.

This system is embedded in an electromagnetic environment which induces an effective transition frequency $\tilde{\omega}= \omega_0 - \delta \omega$ with $\omega_0$ and $\delta \omega$ denoting the bare frequency of the two-level system (TLS) corresponding to Eq.~(\ref{eq:H_S}) and the Lamb shift, respectively.
Due to the broad bandwidth of the reservoir, the TLS cannot be isolated from its surroundings. This is in stark contrast, e.g., to a TLS interacting with a single-mode resonator studied in Ref.~\cite{wallraff}.
Our key idea is to identify the impact of the environment on the level splitting by adding an additional noise source consisting of a resistor $R$ associated with equilibrium voltage fluctuations $\delta V$ as shown in Fig.~\ref{fig:system}. The resistor is directly coupled to the sluice island through a capacitor $C_E$. The advantage is two-fold: this reservoir can be engineered according to a given spectral density and the temperature of the sluice can be tuned by varying the temperature of the resistor inducing a minimal effect on the rest of the circuit including the current detector, e.g., a large Josephson junction in parallel with the sluice~\cite{mottonen06}.\\

The interaction Hamiltonian for the circuit in Fig.~\ref{fig:system} reads  \cite{pekola09, solinasPRB10}
\begin{equation}
{H}_I = - N\,  2e\, \frac{C_{E}}{C_{\Sigma}}\, \delta V
\label{eq:interactionHamiltonian}
\end{equation}
with voltage fluctuations determined by the fluctuation spectrum
\begin{equation}
 S_V(\omega) = \frac{R \hbar \omega}{1+(\omega/\omega_c)^2} \left [ \coth\left( \frac{\hbar \omega}{2 k_B T}\right) + 1\right]
\end{equation}
with the cut-off frequency $\omega_c = (R C_E)^{-1}$. The effective spectral density $J(\omega)$ of this reservoir is obtained from the spectral noise density in Eq.~(\ref{eq:Dfunction}) at low temperatures, namely, $\lim_{T\to 0} D(-\hbar\omega)=2\hbar\, J(\omega)$. We find
\begin{eqnarray}
\lim_{T\to 0} D(-\hbar\omega) &=& 4 e^2 \left(\frac{C_E}{C_{\Sigma}}\right)^2 \, \lim_{T\to 0}S_V(\omega) \nonumber \\
 &\approx& \frac{16 \hbar^2 \pi R}{R_0}  \left(\frac{C_E}{C_{\Sigma}}\right)^2 \frac{\omega}{1+(\omega/\omega_c)^2}\, ,
 \label{eq:S_total}
\end{eqnarray}
where $R_0 = h/e^2$ is the resistance quantum. Hence, in the present circuit the reservoir induces in the relevant low-temperature regime a Drude-type spectral density (\ref{drude})  with the
 dimensionless system--environment coupling constant $\alpha=  8 \pi R/R_0\, (C_E/C_{\Sigma})^2$.
Still within the range of validity of this approximation, we can effectively change the intensity of the Lamb shift by tuning the local temperature.\\

\subsection{Detection of the Lamb shift}

The possibility to control the parameters $J_{L,R}$ and $V_g$ enables the utilization of the optimal detection window for measuring the Lamb shift. We fix the gate voltage to the degeneracy point $N_g=1/2$ and keep both SQUIDs open ($J_{L,R} \neq 0$) but with different energies $J_{L} \neq J_R$~\cite{note_device}.
In this case, the Hamiltonian of the system (\ref{eq:H_S}) reduces to
\begin{equation}
  {H}_S(\varphi) = -J_L \cos \left( \frac{\varphi}{2}- \theta \right)-J_R \cos \left(\frac{\varphi}{2} +\theta \right)\, ,
  \label{eq:H_J}
\end{equation}
where in the restricted Hilbert space spanned by the charge states $|0\rangle$ and $|1\rangle$,
the eigenstates are given by \cite{solinasPRB10}
\begin{eqnarray}
 |g\rangle &=& \frac{1}{\sqrt{2}}\left( |0\rangle + e^{-i \mu} |1\rangle  \right) \nonumber\\
 |e\rangle &=& \frac{1}{\sqrt{2}}\left( |0\rangle - e^{-i \mu} |1\rangle  \right)
 \label{eq:g_e_basis}
\end{eqnarray}
with $\mu = \arctan \left(\frac{J_R-J_L}{J_R+J_L} \tan \frac{\varphi}{2} \right)$. In this basis the Hamiltonian is ${H}_S(\varphi)= (\hbar \omega_0/2)(\EE - \GG)$ with $\hbar\omega_0= \sqrt{J_L^2+J_R^2 + 2 J_L J_R \cos \varphi}$.

 \begin{figure*}[t]
    \begin{center}
    \includegraphics[scale=1]{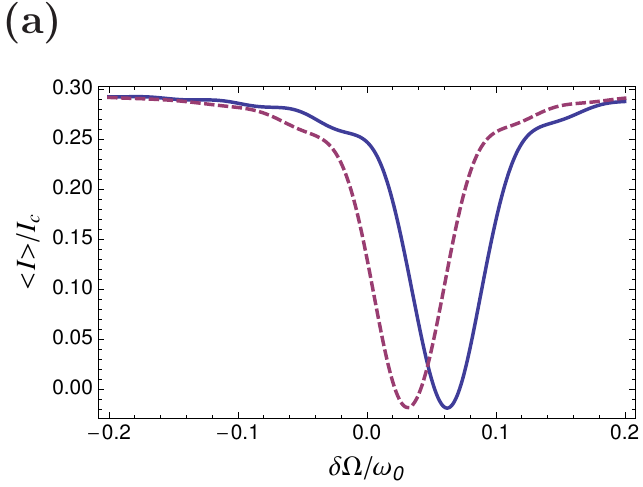}
    \includegraphics[scale=1]{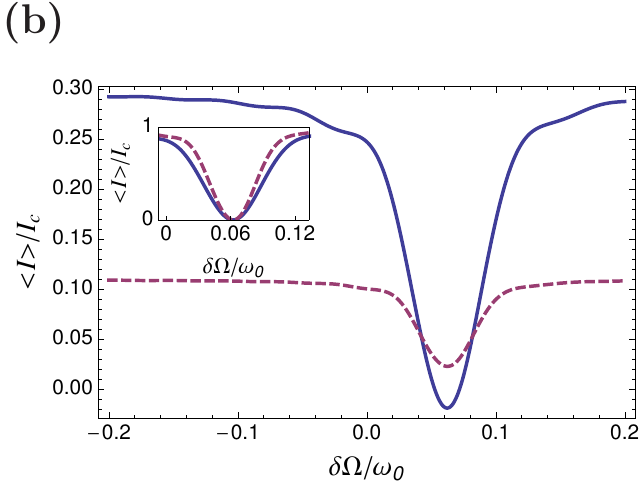}
    \end{center}
    \caption{(Colour online) (a) Average net current in a microwave-driven Cooper pair sluice as a function of the frequency shift $\delta \Omega$. The minimum current is found at $\delta \Omega/\omega_0 = 0.063$ for  $T=50$~mK  (solid line) and at $0.033$ for $T=400$~mK (dashed line). Currents are given in units of the critical current $I_c= 3.14$~nA yielding  $\langle I_{\rm max}\rangle = 0.93$~nA and $\langle I_{\rm min}\rangle = -0.06$~nA. Data are averaged over a total time $t_{\rm fin}=7.6$~ns with the system initialized to the ground state and the microwave drive turned on instantaneously at the beginning of the averaging period.
        (b) Same as in panel (a) but for various values of the phase bias  $\varphi_0 = \pi/2$ (solid curve) and $\varphi_0 = 0.5$ (dashed curve) at $T=50$~mK. The inset shows the normalized shape of the drop for the respective values of $\varphi_0$.
 Other parameters  are $E_C/k_B= 1$~K, $N_g=1/2$, $J_L= 0.1 \,E_C$, $J_R= 0.05\, E_C$, and $\Delta = 0.2$.
\label{fig:numerics}
}
    \label{fig:AverageCurrPhase}
\end{figure*}

The basic idea is to analyse the spectrum of the two-level system, i.e., to measure its transition frequency, by exciting it using a microwave field with tunable frequency. The corresponding Rabi frequency is assumed to be sufficiently high such that the relaxation and dephasing processes do not play any role (see also the discussion at the end of Sec.~\ref{sec:numerics}). Consequently, the interaction with the environment manifests itself only in a Lamb shift $\delta \omega$ which can formally be incorporated in the system Hamiltonian ${H}_S(\varphi)\to {H}_S^{\rm LS}(\varphi)= \hbar/2\,(\omega_0-\delta \omega) (\EE - \GG)$. A static phase bias $\varphi_0$ is supplied by a magnetic flux through the outer superconducting loop and a microwave field coupled inductively to the loop generates a time-periodic phase drive of the form $\delta \varphi(t) = \Delta \cos [ (\omega_0 +\delta \Omega)t]$
 with the off-set $\delta \Omega$ around the bare frequency $\omega_0$ and small amplitude $\Delta\ll\varphi_0$.

Let us study the driven system analytically before carrying out the full numerical calculation in the next section.
To this end, the total Hamiltonian ${H}_{\rm tot}(\varphi)\approx {H}_S^{\rm LS} (\varphi_0) + \delta {H}_{S}$ is expanded up to the first order in $\delta\varphi$ as
\begin{equation}
  \delta {H}_{S}= \left[J_L \sin \left( \frac{\varphi_0}{2}- \theta \right)+J_R \sin \left(\frac{\varphi_0}{2} +\theta \right) \right]\frac{\delta \varphi}{2}\, .
  \label{eq:deltaH_J}
\end{equation}
With the help of the eigenstate basis in Eq.~(\ref{eq:g_e_basis}) it can be expressed as
\begin{equation}
  \delta {H}_{S} = \left [h_{gg} (\EE- \GG)+  h_{ge} \GE + \textrm{h.c.} \right ] \frac{\delta \varphi}{4},
\end{equation}
where $h_{gg}= J_- \sin \mu \cos \frac{\varphi_0}{2}-J_+\cos \mu \sin
  \frac{\varphi_0}{2} $ and  $h_{ge}= i (J_-\cos \mu \cos
  \frac{\varphi_0}{2} +J_+\sin \mu \sin
  \frac{\varphi_0}{2} ) $.

  The procedure is now straightforward. Upon switching to a rotating frame with respect to $(1/2)(\omega_0+\delta\Omega)\,t\cdot(|e\rangle\langle e|-|g\rangle\langle g|)$ in combination with a rotating wave approximation one finds for the probability to reach the excited state at time $t$, provided the system is initially at $t=0$ prepared in the ground state, the known result \cite{cohen}
  \begin{eqnarray}
  P_{ge}(t)&=&\frac{\Delta^2 \, |h_{ge}|^2}{\Delta^2 \, |h_{ge}|^2+16 \hbar^2 (\delta\Omega+\delta\omega)^2}\nonumber\\
   &&\times\sin^2\left[\sqrt{\Delta^2 \, |h_{ge}|^2+16 \hbar^2(\delta\Omega+\delta\omega)^2}\ \frac{t}{8\hbar}\right]\, .\nonumber\\
   \label{eq:Prob_ex}
  \end{eqnarray}
  Hence, in the off-resonance region $|\delta \Omega+ \delta \omega|\gg {\Delta}\,  |h_{ge}|/(4\hbar)$ the system is essentially trapped in the ground state and exhibits low amplitude, i.e., high-frequency oscillations, while in resonance $\delta\Omega+\delta\omega\approx 0$ one has complete transitions with the Rabi frequency $\Delta |h_{ge}|/(8\hbar)$.\\

 %We employ the interaction picture with respect to ${H}_S^{\rm LS}$ and apply the rotating wave %approximation.
%Thus the driving part of the Hamiltonian assumes the form
%\begin{equation}
  %\delta {H}_{S}^{\rm RWA} = \frac{\Delta}{8}  \delta H_{J,ge}\, e^{i t (\delta \Omega +\delta \omega)} %\GE + \textrm{h.c.}
%\end{equation}
%In the off-resonance region $|\delta \Omega+ \delta \omega|\,t\gg {\Delta}\, \delta H_{J,ge}/(4\hbar)$,
 %the driving Hamiltonian $\delta {H}_{S}^{\rm RWA}$  averages effectively to zero and the system remains in %its initially prepared ground state. On the contrary, if the resonance condition
 %$\delta \Omega= - \delta \omega$ is met, $\delta {H}_{S}^{\rm RWA}$ induces Rabi oscillations between  ground and excited state.
The passage from the off-resonance to the resonance region can be observed by monitoring the current through the circuit determined by the current operator $\hat{I} = 2 e \partial_{\varphi_0}{H}_S/\hbar$ since the ground and excited states support currents flowing in opposite directions~\cite{ashhab}.
If the system is initialized in the ground state, a finite current flows through the circuit. As soon as we enter the resonant region, however, transitions are induced such that ground and excited state currents interfere destructively leading ideally to a vanishing net current in the circuit.
The position of this current dip provides the total transition frequency of the system $\omega_0 - \delta \omega$. Its response to the temperature of the engineered environment gives access to the Lamb shift $\delta \omega$.

\section{Numerical results% and estimated value of the Lamb shift in a Cooper pair box
}\label{sec:numerics}

Here, we provide the full numerical simulation of the detection scheme for the Lamb shift introduced in Sec.~\ref{sec:experiment}. The average net current through the sluice is presented in Fig. \ref{fig:numerics} as a function of the frequency off-set $\delta \Omega$.
 For the parameters employed here, we expect a current drop $\Delta I = \,\langle I_{\rm max}\rangle -\langle I_{\rm min}\rangle\, \approx 1$~nA resulting in a detectable signal.
By increasing the  temperature from $0.05$~K to $0.4$~K, the resonance dip is shifted from $0.063\times\omega_0$ [solid curve in Fig.~\ref{fig:numerics}(a)] to $0.033\times\omega_0$ [dashed curve in Fig.~\ref{fig:numerics}(a)] which is sufficiently large for being experimentally observable taking into account the typical current noise at the detector~\cite{mottonen08}. The shape of the dip depends
on both the current operator and the amplitude of the induced Rabi oscillations, which can be varied, for example by tuning $\varphi_0$ [see, for example, Eq. (\ref{eq:Prob_ex})].
In Fig. \ref{fig:numerics}(b), we show current dips for different values of $\varphi_0$.
For decreasing $\varphi_0$, the current drop becomes sharper but the reference current decreases as well yielding a smaller difference  $\Delta I$.
It turns out that for the optimal value $\varphi_0=\pi/2$, the dip is both sharp and deep.
%Notice that the exact evolution of the system in the passage from off-resonant to resonant region depends on $\delta \hat{H}_J$ and then on $\varphi_0$ as well as the current operator.
%These determine the shape and depth of the drop and can influence the accuracy detection of the drop position.
%In Fig. \ref{fig:numerics} (b) we show the behaviour of the current drop for different values of $\varphi_0$.
%When $\varphi_0$ decrease to zero the current drop becomes sharper but at the same time the reference current decreases giving a smaller difference  $\Delta I $.
%It turns out that the the optimal possible values is $\varphi_0=\pi/2$ which produces both a sharp and strong drop.
The experimental observation of the expected temperature dependence of the frequency shift, given in Fig.~\ref{fig:lambshiftexact}, would provide conclusive evidence of  the Lamb shift.

Let us discuss the role of decoherence induced by the environment in our scheme. According to the general theory presented in Sec.~\ref{general}, its impact is  negligible as long as the frequency for Rabi oscillations is sufficiently higher than the decay rate of the coherences $\Gamma_{ee}\sim \omega_0\alpha$ [see Sec.~\ref{TLS}] in the relevant low temperature domain, i.e., $\Delta |h_{ge}| \gg 8   \hbar\omega_0 \alpha$.
To be consistent with this constraint we have assumed a measurement time for the net current of $t_{\rm fin}=\frac{\hbar}{E_C}\cdot 1000\approx 7.6$~ns which for typical device parameters still guarantees a unitary time evolution of the sluice and in particular several tens of Rabi oscillations.

Both the excitation of the system through the perturbation $\delta H_S$ and the measurement of the induced current must be done within $t_{\rm fin}$.
In a realistic experiment, the main challenge in this situation is the synchronization between the two pulses within $t_{\rm fin}$.
If the pulses present an off-set, the average is less accurate since it is performed over a samller number of Rabi oscillations.
However, sequential measurements with different offsets should produce a reliable current average.
An alternative approach is to drive the system continuously, in the case of which there are no issues with timing the pulses and the measurement can be carried out slower. Floquet theory can be a good starting point to study this option in more detail~\cite{Floquet}.

\section{Conclusions}\label{sec:conclusions}

We have presented a careful theoretical treatment of the Lamb shift including both the results for weak-coupling master equations and for a model of an exactly solvable harmonic degree of freedom. Contrarily to the situation of a cavity, where only a single-mode environment is considered, we found it important to analyse the Lamb shift for thermal broadband reservoirs typical for mesoscopic solid state devices. Furthermore, a strategy to measure the Lamb shift in a realistic experimental set-up was discussed in detail supported by numerical simulations. Our results show a shift of about $6\%$ of the bare transition frequency which is considerably larger than the so far measured Lamb shift of up to $1.4\%$ of the qubit transition frequency in a superconducting waveguide resonator \cite{wallraff}. In summary, the experimental realization of our proposal is well motivated.
\vspace{0,5cm}

\section*{Acknowledgements}

The authors thank S. Gasparinetti for stimulating discussions.
Funding was received from the European Community's Seventh Framework Programme under Grant Agreement No. 238345 (GEOMDISS) and from the DFG (SFB/TRR 21). We also acknowledge Academy of Finland and Emil Aaltonen Foundation for financial support.

%\bibliographystyle{plain}
%\bibliography{bibliographiepaper}

\begin{thebibliography}{99}
\bibitem{note_TLS} Note: For a two-level system the counter term in the interaction operator of Eq.~(\ref{eq:Hamiltonian}) cancels out because of $\sigma_z^2=\mathds{1}$.
\bibitem{note_device} Since the parameters $J_{L,R}$ are constant in the model, similar experiments can be performed in devices with asymmetric Josephson junctions. However, use of the SQUIDs and tunable Josephson energies allows to optimize the working point for a given set-up.
\end{thebibliography}

\end{document}